# Significant Light Extraction and Power Efficiency Enhancement of Organic Light Emitting Diodes by Subwavelength Dielectric-Nanomesh Using Large-area Nanoimprint

*Ji Qi, Wei Ding, Qi Zhang, Yuxuan Wang, Hao Chen, and Stephen Y. Chou\**

Prof. S. Y. Chou
Department of Electrical Engineering
Princeton University
Princeton, New Jersey 08544, USA
E-mail: chou@princeton.edu
J. Qi, W. Ding, Q. Zhang, Y. Wang, H. Chen
Department of Electrical Engineering,
Princeton University
Princeton, New Jersey 08544, USA

**Abstract**

To improve the power efficiency of light emitting diodes (OLEDs), we developed a novel OLED structure, termed "Dielectric-Nanomesh OLED" (DNM-OLED), fabricated by large-area nanoimprint lithography (NIL). A dielectric-nanomesh substrate with a subwavelength nanomesh grating patterned into glass releases the photons trapped in ITO and organic layers by guided modes into leaky modes. And the dielectric-nanomesh substrate modifies the entire OLED layers into a corrugated configuration, which further lowers the parasitic resistance and hence the operating voltage. For a red phosphorescent OLED in our demonstration, with the DNM-OLED structure, the light extraction efficiency was enhanced by 1.33 fold and the operating voltage was lowered by 46%, acting together, leading to a 1.8-fold enhancement in power efficiency. Besides red OELDs, the DNM-OLED structure was also shown to be able to work for green and blue OELDs and achieved enhanced light extraction efficiency (1.31 fold for green light, 1.45 fold for blue light) without further structure modification, which demonstrated the light extraction enhancement by the DNM-OLED is broadband.

**Introduction**

One of the central challenges of OLEDs is to enhance the power efficiency (light output over electrical power input), which is further related to the external quantum efficiency (EQE) and the power consumption over the parasitic resistance (hence the operating voltage).   The EQE is related to the intrinsic quantum efficiency (IQE) of the organic light emitting materials and the light extraction efficiency of the OLED device structure. Nowadays, the phosphorescent light emitting molecules have achieved nearly 100% IQE[1, 2], but the light extraction efficiency of the conventional planar OLED device configuration is only about 20% to 30%[3-6]. Hence, the EQE of OLEDs is mainly limited by the light extraction efficiency of the device structure. So far, a number of methods have been proposed for light extraction of OLEDs (e.g., microlens arrays[7-9], high-index substrates[10, 11], light scattering structures[12-18] and microcavity[19, 20]). But most of these common methods only enhance the light extraction and do little help in lowering the parasitic resistance and operating voltage, which make the power efficiency enhancement not very significant. Since the organic materials used in OELD have low carrier motilities and are difficult for carrier injection compared to the semiconductor materials used in solid state lights, the power consumption over the parasitic resistance of the organic materials take large amount of the total power consumption and considerably affect the power efficiency. As for lowering the parasitic resistance (hence the operating voltage) of OLEDs, nowadays the typical way is to develop new hole or electron transport materials with high carrier motilities and new electrode materials for facilitating carrier injection[21,22]. OLED device structures that both enhance the light extraction and reduce the parasitic resistance without introducing new materials have been rarely reported.

Microstructure or nanostructure patterned substrate is one of the promising ways for enhancing the light extraction efficiency of LEDs[13-18]. Previously, the substrates patterned with micro-mesh[12], random nano-patterns[16] and periodic nano-pillars[17-18] have been studied for light extraction of OLEDs. Among those various patterns, the micro- or nano-mesh draw great attention of researchers, due to the relatively easy fabrication and good scattering efficiency. The researchers Yiru Sun *et al.* first proposed to use mesh patterned substrates to extract the trapped light in OLEDs[12]. The OLED devices in their studies, with low-index grids embedded in the organic layers between two electrodes, achieved $32\pm6\%$ light extraction enhancement compared to the ITO controls. But for the configuration of their device, the dielectric low-index grids embedded in the organic layers reduced the active area at the interface between the organic layers and electrodes. As a result, increased parasitic resistance and operating voltage of their devices should be anticipated, which causes negative influence on improving the power efficiency. Hence, a new device structure that utilizes the micro- or nano-mesh patterns for both light extraction enhancement and parasitic resistance reduction is highly desired. Moreover, many of patterned substrates nowadays are fabricated by laser interference lithography[25], electron beam lithography (EBL)[26] or focused ion beam (FIB) lithography[27], which are complicated and time-consuming, and difficult to be applied in real large-area manufacture process with high throughput. Hence, easy fabrication processes are highly desired for manufacturing micro- or nano- patterned substrates in high throughput.

In this report, we developed a new substrate patterned with nanomesh, termed dielectric-nanomesh substrate, which can be easily fabricated by large-area nanoimprint lithography. And by directly depositing OLEDs materials on top of the dielectric-nanomesh substrate, we fabricated a novel corrugated OLED structure, Dielectric-Nanomesh OLED (DNM-OLED), which experimentally proved to be able to achieve significant broadband light extraction enhancement and lower the parasitic resistance (i.e., lower the operating voltage), and hence lead to a considerable increase in power efficiency.

**DNM-OLED structure, design and principle.** The key component in DNM-OLED is the dielectric-nanomesh substrate which is a glass substrate with a subwavelength nanomesh grating etching into it (Fig. 1b). On top of the dielectric-nanomesh substrate, the (indium-tin-oxide) ITO front transparent electrode, the organic light emitting active layers and aluminum (Al) back electrodes are sequentially deposited to form the final DNM-OLED structure. The morphology of the dielectric-nanomesh substrate also modifies the materials deposited on top of it, and therefore the entire OLED device has a corrugated nanomesh configuration (Fig. 1a). In DNM-OLED operation, holes and electrons are injected from ITO and Al respectively into the light emitting materials to generate photons and light comes out through the transparent dielectric-nanomesh substrate. In an optimized red DNM-OLED fabricated, the dielectric nanomesh on the glass substrate has a 400nm-pitch square lattice with periodic square hole arrays. The line width of the nanomesh grid is 75nm and the optimized groove depth is 40nn (discussed later). On top of the dielectric-nanomesh, the front transparent electrode is 140nm thick ITO; the back electrode is 0.5nm thick LiF and 100nm thick Al; and in between the light emitting materials consist of 40nm thick N,N'-Bis(naphthalen-1-yl)-N,N'-bis(phenyl)benzidine (NPB) as the hole transport material, 30nm thick host material 4,4'-Bis(N-carbazolyl)-1,1'-biphenyl (CBP) doped with 6 wt% high efficient red emitting phosphorescent guest molecules Bis(1-phenylisoquinoline)(acetylacetonate)iridium(III) [Ir(piq)2acac], and 50nm thick 2,2',2"-(1,3,5-Benzinetriyl)-tris(1-phenyl-1-H-benzimidazole) [TPBI] as the electron transport material.

The dielectric-nanomesh substrate plays an important role in increasing light extraction efficiency and lowering the operating voltage. For a conventional OLED with the planar structure, only about 20% of the photons with lateral wave vectors smaller than the amplitude of the wave vector in air (i.e., $k_0 = \frac{2\pi}{\lambda_0}$, where $\lambda_0$ is the wavelength in air) can be out-coupled into air, and other 80% with large lateral wave vectors ($k_0 < |\mathbf{k_{trap}}| < k_n$, where $\mathbf{k_{trap}}$ is the lateral wave vector of trapped photon and $k_n = \frac{n2\pi}{\lambda_0}$ is the amplitude of wave vector in ITO and organic active layers) are trapped in OLED layers by substrate, wave guide and SPP modes. But for a DNM-OLED, the trapped photons will be coherently scattered by the periodic nanomesh grids on the substrate and Al back electrode, and receive an additional lattice momentum ($\mathbf{\Lambda} = \frac{n_x 2\pi}{a}\mathbf{i} + \frac{n_y 2\pi}{a}\mathbf{j}$, where a is the pitch of nanomesh) so that they are able to be coupled to the leaky modes and released into air if satisfying the coupling condition, i.e., $|\mathbf{k_{trap}} \pm \mathbf{\Lambda}| < k_0$. Considering the refractive index (n) of the ITO and organic layers is around 1.8, the 400nm period of the dielectric-nanomesh is found well chosen. By calculation, in a broad range of wavelength from 320nm to 800nm, all the trapped states ($k_0<|\boldsymbol{k}_{\mathbf{trap}}|<k_\mathrm{n}$) can satisfy the coupling condition. Hence, the DNM-OLED is perfect for broadband light extraction enhancement and later we will see that DNM-OLED structure can achieve increasing light extraction for all red, green and blue OLEDs. Besides enhancing light extraction, as shown later, the DNM-OLED is able to reduce the parasitic resistance. The corrugated nanomesh configuration of the DNM-OLED increases the active interface area between the electrodes and organic layers and partially reduce the thickness between two electrodes, which was experimentally demonstrated to be able to reduce the parasitic resistance and hence the operating voltage.

**Fabrication**

The fabrication of the DNM-OLED starts from fabricating the dielectric-nanomesh substrates. The fabrication process is schematically illustrated in Fig. 2. A chromium (Cr) nanomesh was first fabricated on a 0.5mm thick glass substrate by NIL and the deposition and lift-off of 10nm thick Cr. Then the final dielectric-nanomesh substrate was achieved by etching the glass substrate using Reactive-ion etching (RIE) with the Cr nanomesh mask, followed by removal of the Cr mask. The etching depth is 40nm. And as shown in Fig. 3a, in the final step, the red emitting DNM-OLED was fabricated by sequentially depositing 140nm ITO layer by e-beam evaporation, 40nm thick NPB, 30nm thick CBP doped with 6wt% Ir(piq)₂acac, 50nm TPBI and 0.5nm/100nm thick LiF/Al by thermal evaporation under high vacuum (<10⁻⁷ torr). The light emitting area of a DNM-OLED device is 3mm by 3mm, which is defined by a shadow mask during the evaporation of Al back electrode.

The Scanning electron microscopy (SEM) shows that the dielectric-nanomesh in the substrate indeed has a 400nm pitch, 75nm line width and 40nm groove depth as we designed, and excellent nanopattern uniformity over large area (Fig. 3c: left). And the SEM of the cross-section of the final DNM-OLED (Fig. 3c: right) shows that the organic active layers deposited on the ITO anode are very smooth without have any observable pinholes. And as we expected, the topology of the dielectric-nanomesh substrate indeed modifies the materials deposited on top of it into a corrugated configuration although the modification gets weaker for layers further away from the substrate.

For comparison, reference OLEDs, "ITO-OLEDs" were also fabricated, which are the same OLEDs as the DNM-OLEDs except replacing the dielectric-nanomesh substrates with planar glass substrates.

**Results and discussion**

**Electroluminescence, EQE and light extraction enhancement.** The total front surface electroluminescence (EL) spectra of the red emitting DNM-OLED and ITO-OLED (Fig. 4a) were measured using an integrated sphere (Labsphere LMS-100) connected to a spectrometer (Horiba Jobin Yvon). During the measurement, the devices are attached on a stage holder in the integrated sphere and both the side walls and the back surface of the devices were fully covered by black tapes to ensure the light emission only from the OLEDs' front surfaces. The measured EL spectra show that in the entire emitting range (550nm to 800nm), the EL intensity of DNM-OLED is much higher than that of ITO-OLED. Compared with the ITO-OELD, at the emission peak (628nm), the DNM-OLED shows a 1.25 fold EL enhancement factor (i.e. the ratio of the spectrum of DNM-OLED to ITO-OLED) and by averaging the EL enhancement factor over the entire emission range, the DNM-OLED shows a 1.44 fold average EL enhancement factor. The measurements clearly show that EL enhancement by the DNM-OLED structure is broadband, although the EL enhancement factors measured at longer wavelengths are a litter higher than that measured at shorter wavelengths (within 20% max. variation) in the emission range. The broadband EL enhancement is because that the dielectric-nanomesh is a broadband light extracting structure as discussed above. Actually the EL enhancement bandwidth of the DNM-OLED should be much wider than what is showed in Fig. 4a. For this particular red emitting DNM-OLED, the enhancement bandwidth is limited by the emission bandwidth of the red light emitting material. Later, it will be shown that by applying the DNM-OLED structure to green and blue light emitting materials, the DNM-OLED has demonstrated to achieve light extraction over the entire visible wavelength range.

The EQE as a function of injection current density of the DNM-OLEDs and ITO-OLEDs (Fig. 4b) were obtained by converting the measured EL spectra to power spectral density (W/nm) at a given current density using a calibrated lamp standard (Labsphere AUX-100). The measured

EQE shows that in the current density range from 1mA/cm² to 100mA/cm², the DNM-OELD exhibits a maximum EQE of 16% (at current density <5mA/cm²), which is 1.33 fold higher than that of the ITO-OELD (a maximum EQE of 12%). And the EQE of 16%, to our best knowledge, is the highest achieved for red emitting OELDs using CBP as the host material in the emission layer. Considering $EQE = IQE \times \eta_{ex.}$ where $\eta_{ex.}$ is the light extraction efficiency, and we assume the IQE of the light emitting materials is not affected by the different OELD structures, hence the light extraction efficiency enhancement should be the same as the EQE enhancement. Therefore, based on the measured EQE, the light extraction enhancement for this red emitting OLEDs by DNM-OLED structure is 1.33 fold.

**Current density-voltage characteristic, lowered operating voltage and power efficiency enhancement.** The current density vs. operating voltage (J-V) characteristics were measured for both the DNM-OLED and the ITO-OLED (Fig. 4c) using a source meter (HP 4145B). Compared with the ITO-OLED, the DNM-OLED shows much larger current increasing slope and significantly lowered operating voltage at a given current density. The operating voltage has a maximum reduction by 46% from 12V to 6.5 V at 6mA/cm². The lowered operating voltage means that the parasitic resistance of the OLED is lowered by the DNM-OLED structure and it could be attributed to two possible mechanisms. Firstly, the corrugated configuration of the DNM-OLED increases the active interface between the organic active layers and electrodes, hence lowering the parasitic resistance and increasing the total current going through the device at a given operating voltage. For this particular DNM-OELD, by calculation using the geometrical parameters directly measured from the SEM of the device cross-section (Fig. 3c), the total interface area between ITO and organic active layers increases by 15% compared with a planar interface. Since the total current is linearly proportional to the interface area, hence the 15% increased interface area cannot solely explain the significantly lowered operating voltage at a given current density or increased current density at a given operating voltage as shown in Fig. 4c. Another mechanism for the lowered parasitic resistance is that the partial reduction of the distance between the anode (ITO) and cathode (Al) resulting from the corrugated device configuration enhances the electric filed intensity in the organic active layers [18]. Because in organic materials the current density as a function of the electric field follows the power law[18, 28], hence the enhanced electric field intensity could lead to a considerable increase in current density at a given operating voltage or reduction in operating voltage at a given current density.

Using the measured J-V characteristics and EL spectra at different injection current densities, we obtained the power efficiency as a function injection current density of the DNM-OELD and the ITO-OELD (Fig. 4d) by integrating the EL spectra with the human eye's luminosity function and dividing by the input electrical power. Compared with the ITO-OLED, the maximum power efficiency of DNM-OELD at 1mA/cm$^2$ shows 1.8-fold enhancement increasing from 4.2lm/W to 7.7lm/W. Because the power efficiency of OLEDs is the ratio of outcoupled luminous power to the input electrical power, therefore the significant power efficiency enhancement by DNM-OLED is a combined effect of both enhanced light extraction and lowered parasitic resistance.

**Emission profiles and angular dependence of EL spectra and luminance intensity.** The normalized angular dependence of EL spectra for the DNM-OLED and ITO-OLED were measured at 1mA/cm$^2$ using a rotation stage, collimation lens connected to a spectrometer with an optical fiber and were shown in Fig. 5a and 5b. The measured angular dependence of EL spectra clearly show that in the emission angle range from 0 to 70 degree, the EL spectra of the ITO-OELD are nearly independent of emission angles, while EL spectra of the DNM-OLED show strong angular dependence and the emission peaks have blue shifts as the emission angles increase. This EL emission peak shift of the DNM-OLED is because the wavelength of photons that achieve enhanced extraction is angular dependent and could be explained by the coupling equation, i.e. $k_0 \sin\theta = |\boldsymbol{k_{trap}} \pm \boldsymbol{\Lambda}|$ where $\theta$ is the emission angle, which further confirms the light extraction enhancement by DNM-OLED is due to the coupling between trapped modes and leaky modes through lattice momentum as discussed before. When the emission angle increases from 0 to 40 degree, the EL emission peak of DNM-OELD shifts from 650nm to 605nm. While for the emission angle larger than 40 degree, because the wavelength achieved enhanced extraction is out of the bandwidth of the emission spectrum, hence the EL spectra still show the intrinsic emission peak of the light emitting material.

And the 2D far filed emission profiles of the ITO-OLED and the DNM-OLED were obtained by taking images of the operating OLEDs in normal direction using CCD (Fig. 5c). It can be clearly observed that the ITO-OELD shows an isotropic emission distribution in all directions, whereas the DNM-OELD shows an emission pattern with a four-fold symmetry that exhibits stronger emission intensities along these two periodic directions of the dielectric-nanomesh which agrees well with diffraction pattern of a 2D grating with square lattice and square hole. Along the periodic direction of the dielectric-nanomesh, by integrating the EL spectra over the wavelength with the luminosity function as the weight the normalized luminance vs. emission angle of ITO-OELD and DNM-OELD was obtained (Fig. 5d). As expected, the ITO-OELD's luminance angular distribution is the conventional Lambertian type. But for the DNM-OELD, the strongest luminance intensity is observed at the emission angle of 30 degree, which can be explained by the matching between the wavelength of the photons that achieved enhanced light extraction at emission angle of 30 degree and the emission peak wavelength of the light emitting materials.

**Groove depth dependent light extraction efficiency enhancement.** The principle of light extraction enhancement by the DNM-OLED structure, as discussed above, is mainly due to the coherent photon scattering by the nanomesh grids on the dielectric-nanomesh substrate. Hence, we expect the groove depth of the dielectric-nanomesh on the substrate can effectively affect the light extraction enhancement by DNM-OLED since the scattering efficiency of the nanomesh grids should be considerably affected by the groove depth. And the deeper the groove depth is, the stronger scattering the dielectric-nanomesh should achieve. We fabricated 5 DNM-OELDs with the same device structure except different groove depths (15nm, 25nm, 40nm, 55nm and 70nm). The relative light extraction efficiency as a function of the groove depth was obtained by measuring the maximum EQE of the DNM-OLED at a given groove depth and dividing by the maximum EQE of the ITO-OLED (Fig. 6). One can see that the highest relative light extraction efficiency is achieved at the groove depth of 40nm and what is more interesting to note is that the light extraction efficiency of DNM-OLED is even lower than the ITO-OLED for deep groove depth of 70nm. The possible reason for this groove-depth dependence of light extraction enhancement of the DNM-OELD is that for groove depths smaller than 40nm, the nanomesh grids cannot effectively scatter the light, whereas for very large groove depth (i.e., 70nm in this case) the organic active layers cannot sustain such large corrugation, which makes the electrical properties degrade and cancel the enhanced light extraction.

**Extend DNM-OELD structure to green and blue emitting OELDs.** To further experimentally demonstrate that the DNM-OLED can achieve broadband light extraction enhancement in the entire visible light range, we fabricated green and blue emitting DNM-OELDs and the corresponding ITO-OELDs and measured the EQE as a function of the injection current (Fig. 7). The green and blue emitting DNM-OLEDs have the same device structure as the red emitting DNM-OLEDs discussed above except the light emitting layers. The light emitting layer of the green emitting OLED is NPB (40nm)/ TCTA (20nm)/CBP: Ir(piq)$_3$ (6wt%)(30nm)/TPBI (50nm). Tris(4-carbazoyl-9-ylphenyl)amine (TCTA) is the hole injection material and Tris(1-phenylisoquinoline)iridium(III) [Ir(piq)3] is the high efficient phosphorescent dopants that emit green light (emission peak: 520nm~550nm). The light emitting layer for blue emitting OLED is NPB (40nm)/mCP (20nm)/SPPO1:FIrpic (10wt%) (30nm)/SPPO1 (40nm), where 1,3-Di(9H-carbazol-9-yl)benzene (mCP) is the exciton blocking material and 9,9-Spirobifluoren-2-yl-diphenylphosphine oxide (SPPO1) is the host material for high efficient blue emitting phosphorescent dopants Bis[2-(4,6-difluorophenyl)pyridinato-C2,N] -(picolinato)iridium(III) [FIrpic] (emission peak: 475nm~525nm ) and electron transport material.

The measured EQE vs. injection current density shows that, compared with the ITO-OLED, the maximum EQE of the DNM-OELD (at 1mA/cm$^2$) exhibits 1.31-fold enhancement increasing from 26% to 34% for green light (Fig. 7a) and 1.45-fold enhancement increasing from 22% to 32% for blue light (Fig. 7b). Combined with the result of EQE enhancement by DNM-OLEDs for red emitting OLEDs as shown above, it has been experimentally demonstrated that the DNM-OLED structure is well designed for broadband light extraction enhancement covering the whole visible light range.

**Conclusion**

In summary, we developed a novel OLED structure, DNM-OLED, containing a newly designed substrate patterned with dielectric-nanomesh. Experimentally, the DNM-OLEDs showed considerably improved performances. On the one hand, the DNM-OLEDs exhibited broadband light extraction efficiency enhancement. Compared with the ITO-OLEDs, the DNM-OELDs showed maximum EQE increased from 12% to 16 for red light (1.33 fold enhancement factor), from 22% to 32% for blue light (1.45 fold enhancement factor) and from 26% to 34% for green light (1.31 fold enhancement factor). The broadband light extraction enhancement is because the well-designed dielectric-nanomesh substrate can lead to a broadband coupling between the trapped modes and the leaky modes by coherent scattering. On the other hand, the DNM-OLEDs have demonstrated to be able to reduce the parasitic resistance and lower the operating voltage and hence significantly increase the power efficiency. Using a red DNM-OLED as an example, compared with the ITO-OELD, the operating voltage and maximum power efficiency of the DNM-OLED is reduced by 46% at most and increased by 1.8 fold respectively. This reduction of operating voltage by the DNM-OLED is due to the increased area at the interface between the electrodes and the organic active layers and the reduction of the thickness between two electrodes as a result of the corrugated configuration. Moreover, another advantage of the DNM-OLEDs is the easy and scalable fabrication by nanoimprint lithography. With improvements in light emitting materials and optimization of structure design, the above DNM-OLEDs' performances can be further improved. The designs, fabrications, and findings are applicable to the LEDs in other materials (organic or inorganic) and on other thin substrates (plastics or glasses), thus opening up new avenues in developing high efficient OLED lighting and display.

**Figures**

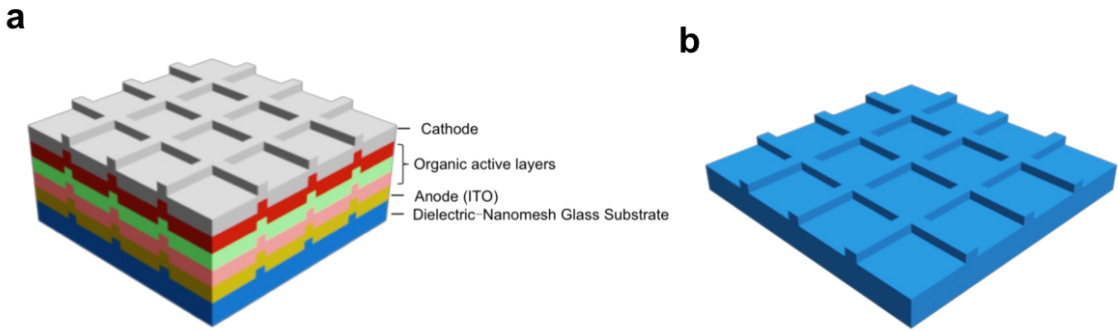

**Figure 1. Dielectric-Nanomesh Organic Light Emitting Diode (DNM-OLED).** (a) The layer structure schematic: a bottom dielectric-nanomesh substrate; anode, organic active layers and cathode deposited on top of it forming a multilayer-corrugated structure; (b) structure schematic of the dielectric nanomesh substrate: a glass substrate with a subwavelength nanomesh etched into it.

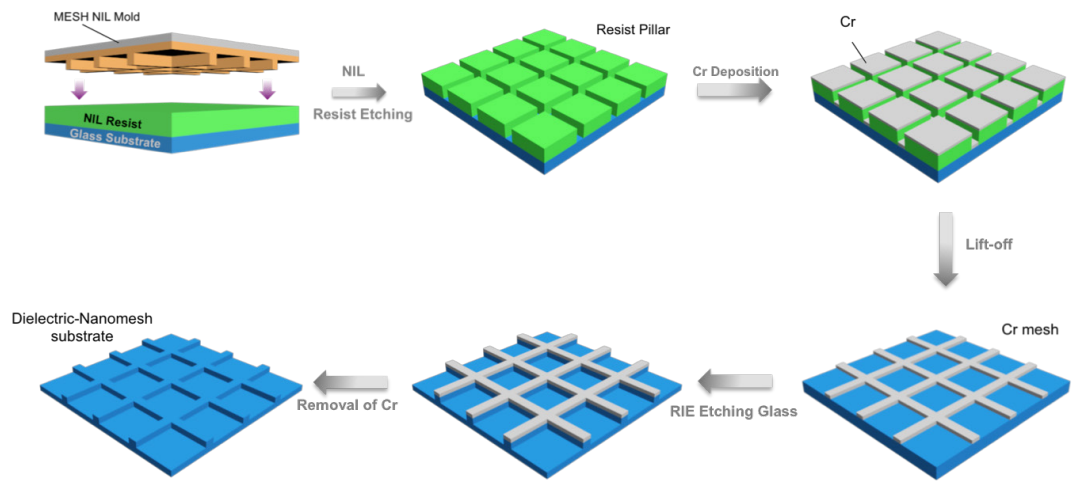

**Figure 2. Fabrication of Dielectric-Nanomesh substrate.** fabrication process: fabrication of Cr nanomesh on a glass substrate by nanoimprint, Cr deposition and lift-off, RIE etching of glass substrate with Cr nanomesh as mask, removal of Cr .

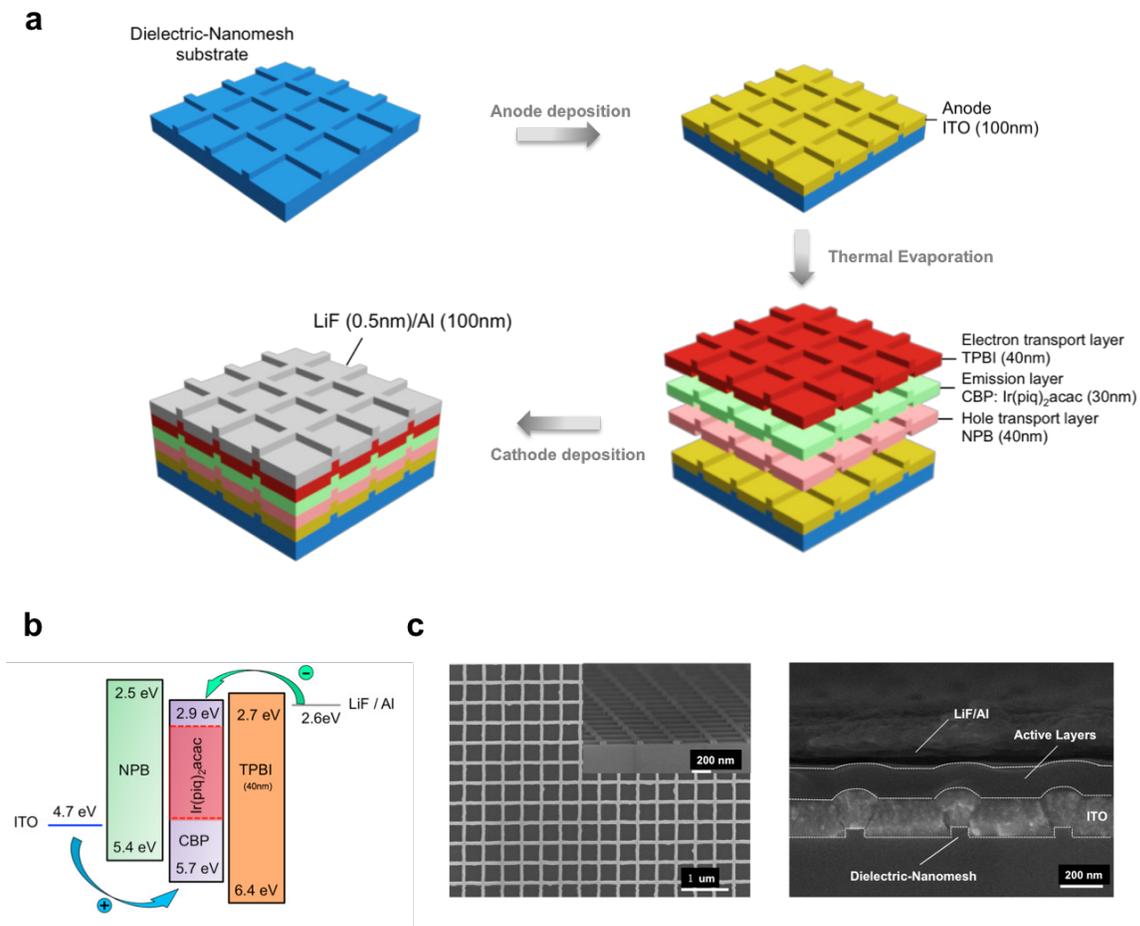

**Figure 3. Fabrication of the red emitting DNM-OLED.** (a) schematic of the fabrication process, including anode (ITO), organic active layers (NPB/CBP: Ir(piq)$_2$acac 6 wt%/TPBI) and cathode (LiF/Al) deposition on the fabricated dielectric-nanomesh substrate; (b) energy band diagram; (c) (left) scanning electron micrograph (SEM) of Dielectric-Nanomesh substrate: 400nm pitch, 75nm line width and 40nm groove depth; (right) cross-sectional SEM of the red emitting DNM-OELD (white dashed lines indicate the interfaces): organic layers deposited on the corrugated ITO electrode exhibit very smooth morphology.

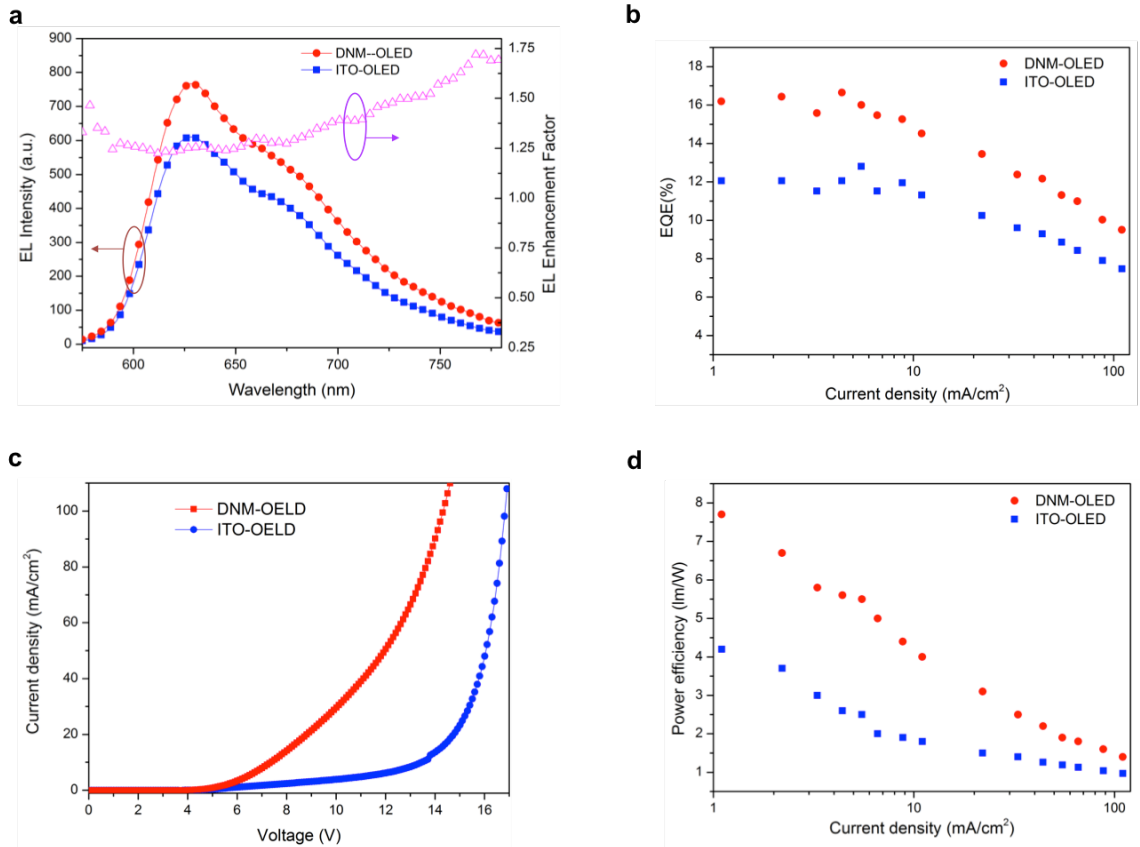

**Figure 4. Measured electro-luminance (EL), EQE, J-V and power efficiency of DNM-OLEDs and ITO-LEDs.** (a) Total front-surface EL/enhancement spectrum; (b) EQE vs. current density; (c) current density vs. driving voltage; (d) power efficiency vs. current density. Compared with ITO-OLEDs, DNM-OLEDs show 1.44-fold average EL enhancement, 1.33-fold max. EQE enhancement, lowered driving voltage by 46% (at 6mA/m$^2$) and 1.8-fold max. power efficiency enhancement.

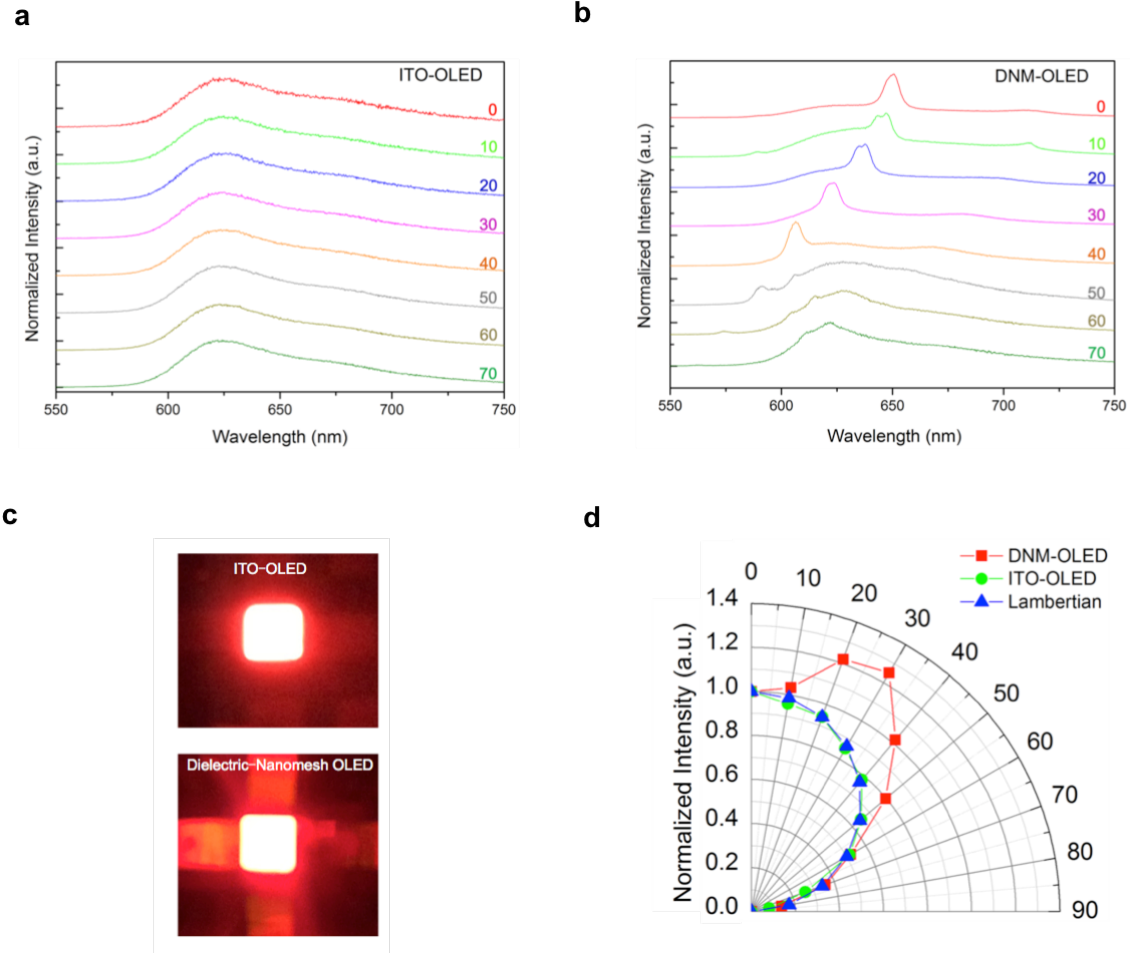

**Figure 5. Angular-dependent EL spectra, luminance and emission profiles of DNM-OLEDs and ITO-LEDs.** (a) normalized EL spectra of ITO-OLEDs (b) normalized spectra of DNM-OLEDs (c) photos of working ITO-OLEDs (upper) and DNM-OLEDs (bottom) indicating the 2D far-field intensity profiles (d) normalized luminance vs. emission angle of ITO-OLEDs and DNM-OLEDs (along the periodic direction of nanomesh).

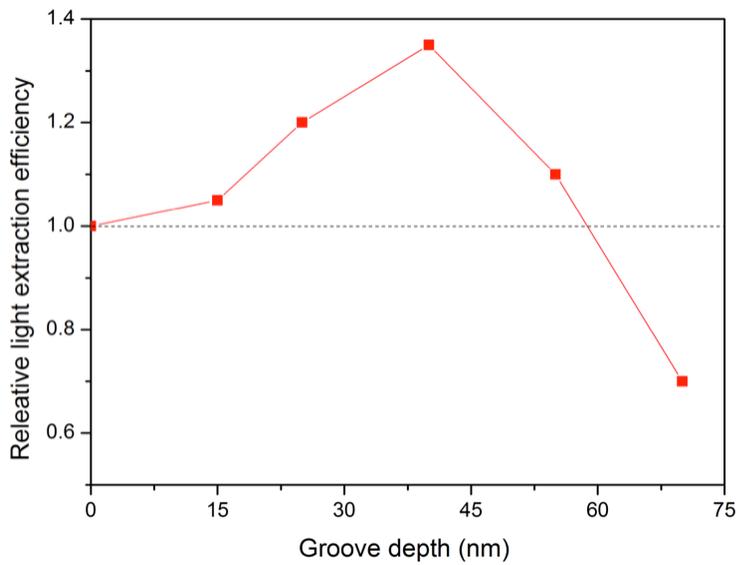

**Figure 6. Light extraction efficiency of DNM-OLEDs with respect to the ITO-OLEDs as a function of the groove depth of the dielectric-nanomesh substrate.**

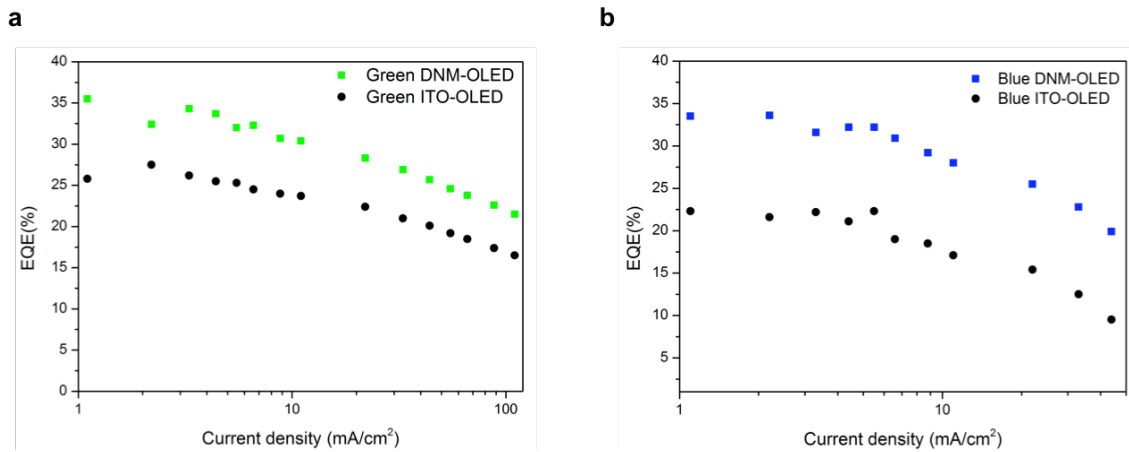

**Figure 7. Measured EQE of Green and Blue DNM-OLEDs.** EQE vs. current density of green (a) and blue (b) DNM-OLEDs. Compared with the ITO-OLLEDs, the maximum EQE of the DNM-OELDs exhibit 1.31-fold enhancement increasing from 26% to 34% for green light and 1.45-fold enhancement increasing from 22% to 32% for blue light.